\documentstyle [epsf]{article}
\begin{document}
\def\ba{\begin{eqnarray}}
\def\ea{\end{eqnarray}}
\def\be{\begin{equation}}
\def\ee{\end{equation}}
\def\({\left(}
\def\){\right)}
\def\[{\left[}
\def\]{\right]}
\def\lagrange {{\cal L}}
\def\del {\nabla}
\def\d {\partial}
\def\ppsi {\mbox{\boldmath $\psi$}}
\def\cchi {\mbox{\boldmath $\chi$}}
\def\SSigma {\mbox{\boldmath $\Sigma$}}
\def\makespace {\qquad\qquad}
\def\Tr{{\rm Tr}}
\def\Im{{\rm Im}}
\def\simlt{ \hbox{\lower 1.ex\hbox{$\stackrel{\mbox{$<$}}{\sim}$}} }
\def\simgt{ \hbox{\lower 1.ex\hbox{$\stackrel{\mbox{$>$}}{\sim}$}} }
\def\<{\left\langle}
\def\>{\right\rangle}
\newcommand{\labeq}[1] {\label{eq:#1}}
\newcommand{\eqn}[1] {(\ref{eq:#1})}
\newcommand{\labfig}[1] {\label{fig:#1}}
\newcommand{\fig}[1] {\ref{fig:#1}}
\newcommand\bi{\bibitem}
\newcommand\bigdot[1] {\stackrel{\mbox{{\huge .}}}{#1}}
\newcommand\derv[1] {{{\rm d}\over{\rm d}#1}}
\newcommand\dee {{\,\rm d}}
\newcommand\Ad[1] {{\rm Ad}\[#1\]}

\catcode`\@=11 \catcode`\:=11
\newcommand\setceqc[2] {
\edef \tempzz{\csname c@#1\endcsname=\csname c@#2\endcsname}\tempzz
}
\catcode`\@=12 \catcode`\:=12    


\begin{center}
\vspace{2cm}
{\LARGE Cosmic Texture from a Broken Global SU(3) Symmetry}\\
\vspace{1cm}
{\large Christopher Barnes\footnote{email: {\it
barnes@puhep1.princeton.edu}}}\\
{\it Princeton University}\\
{\it Joseph Henry Laboratories, PO Box 708}\\
{\it Princeton, NJ 08544, USA}\\
\vspace{1cm}
{\large Neil Turok\footnote{email: {\it N.G.Turok@amtp.cam.ac.uk}}}\\
{\it DAMTP, Silver Street}\\
{\it Cambridge, CB3 9EW, UK}\\
\end{center}

\vspace{1cm} \centerline{\large \bf Abstract} We investigate the
observable consequences of creating cosmic texture by breaking a
global SU(3) symmetry, rather than the SU(2) case which is generally
studied.  To this end, we study the nonlinear sigma model for a
totally broken SU(3) symmetry, and develop a technique for numerically
solving the classical field equations.  This technique is applied in a
cosmological context: the energy of the collapsing SU(3) texture field
is used as a gravitational source for the production of
perturbations in the
primordial fluids of the early universe.  From these calculations, we
make predictions about the appearance of the anisotropies in the
cosmic microwave background radiation (CMBR) which would be present if the
large scale structure of the universe was gravitationally seeded by
the collapse of SU(3) textures.
\vspace{.2cm}

\section{Introduction}
Over the last few years, there has been a great deal of research on
the effects of topological defects in the the early universe.  In
particular, one class of topological defect, textures, has emerged as
an interesting canidate for creating the seeds of large scale
structure in the early universe~\cite{first_text}.
In three spatial dimensions, textures can be created in any symmetry
breaking scheme where the vacuum manifold has a 
nontrivial third homotopy group,
$\pi_3$.  This situation is quite generic - for example if any nonabelian
Lie group is completely broken, $\pi_3$ is nontrivial.
There are a multitude of symmetry breaking 
schemes which could produce textures (a general classification has been given 
in \cite{bryan}). But so far the only case which has been studied
in any detail has been the simplest case, where the vacuum manifold is
SU(2)${}\simeq{}\rm S^3$.  An
accurate
nonlinear sigma model code has been developed for SU(2)
textures
al.\cite{bigso3}, and 
it has been possible to extract detailed predictions of
the power spectrum of inhomogeneities 
and nongaussianity of the CMBR anisotropies
\cite{bigso3}, \cite{ntapjlett}.

In the light of such investigations, it is natural 
to ask whether the primary results from these simulations are particular to
SU(2), or whether they are more general, pertaining to many or all field
theories which would create texture.  To answer this question, one
must study the effects of textures that come from at least one other group.
All simple  Lie groups have a nontrivial $\pi_3$
so in principle one might choose any one of them. However, 
simplicity dictates
SU(3) as the obvious choice. In this paper we shall
study the effects of textures coming from a broken global SU(3) 
symmetry upon the CMBR anisotropies and 
density fluctuations of the early universe.
A complementary investigation of more general
texture theories has been recently conducted by Sornberger et. al. 
\cite{sornborger}.

There is also some motivation from particle physics to study the
effects of a totally broken global SU(3) symmetry.  Joyce and
Turok,~\cite{famsymm}, have pointed out that if the family symmetry
among the quarks and leptons is broken by the Higgs mechanism, then we
must have Higgs fields which totally break some three dimensional
representation of a simple Lie Group.  The only simple Lie groups
which have three dimensional representations are SU(2) and SU(3).
Joyce and Turok argue that SU(2) does not work well, as it does not
easily break to make one quark family much more massive than the
others, whereas in SU(3) this occurs naturally.  A
gauged SU(3) family symmetry would be anomolous, so such an SU(3)
family symmetry must be global.  When this symmetry is broken, 
the Goldstone modes of the Higgs fields
would produce cosmic texture. In particle physics terms, 
the point of the present paper is to explore the cosmological 
consequences of a global
SU(3) family symmetry broken at the GUT scale. 

In this paper, a technique for studying textures from a broken global
SU(3) symmetry is developed and applied, with the goal of making 
predictions about fluctuations in the cosmic microwave background.  
In section~\ref{formulation}, we state the problem and derive the 
relevant nonlinear
sigma model equations
~\cite{famsymm}.  In 
section~\ref{numalg}, we develop a finite difference algorithm based upon
these equations, and in section~\ref{testcode} we
describe tests of the stability and accuracy of this algorithm.  In section
\ref{results} we present some results about the scaling behavior of this 
algorithm, and the appearance of the CMBR fluctuations if the primordial
density fluctuations of the universe come from this type of texture.

We find that the cosmological consequences of
textures coming from a totally broken SU(3) symmetry are in 
most respects very similar
to those from SU(2) textures. We present some preliminary evidence
of differences in the shape of the CMBR anisotropy power spectrum,
but more accurate techniques 
will be needed to fully resolve these. 

\section{Formulation of the problem}
\label{formulation}

We wish to totally break the fundamental representation of SU(3) via
the Higgs mechanism.  In~\cite{famsymm}, Joyce and Turok describe this
process in detail. In order to do this, our Higgs field must consist
of two complex triplets, $\SSigma^1,\SSigma^2\in {\rm C}^3$, which
have a ``Mexican hat'' potential whose minimum is topologically 
equivalent to SU(3).
Without loss of generality, one can choose these fields so that the
minimum of the potential is described by: \be
{\SSigma^1}^\dagger\SSigma^1 = v_1^2 , \qquad
{\SSigma^2}^\dagger\SSigma^2 = v_2^2 , \qquad
{\SSigma^1}^\dagger\SSigma^2 = 0\ .  \labeq{manifold} \ee Here
$v_1,v_2$ are the vacuum expectation values of $\SSigma^1,\SSigma^2$,
and are not necessarily equal to one another. (See equation~(62) of
\cite{famsymm}.) Any particular set of $\SSigma^1,\SSigma^2$
satisfying \eqn{manifold} clearly define a unique member of SU(3), for
if we define \be \ppsi = {\SSigma^1\over v_1}, \qquad \cchi =
{\SSigma^2\over v_2}, \labeq{defpsichi} \ee then we can immediately
construct an SU(3) group element, in the fundamental representation, \be U =
\(
\begin{array}{ccc}
|&|&| \\
\ppsi^\dagger\times\cchi^\dagger&\ppsi&\cchi \\
|&|&|
\end {array}
  \) \ .
\labeq{Udef}
\ee

Note that the vacuum manifold is only identical to SU(3) geometrically
if $v_1=v_2$. Otherwise it represents a `stretched' version of SU(3).

 We wish to construct a nonlinear sigma model which will describe the classical
 dynamics of these fields, assuming that they are restricted to this 
minimum.

The first thing we need are the classical field equations. The 
Lagrangian density for this system is 
\ba
\lagrange &=& v_1^2\del_\mu\ppsi^\dagger\del^\mu\ppsi +
              v_2^2\del_\mu\cchi^\dagger\del^\mu\cchi +  \nonumber \\ 
	  & & \qquad \lambda_\psi(\ppsi^\dagger\ppsi -1) +
	      \lambda_\chi(\cchi^\dagger\cchi -1) +
	   \lambda_{\psi\chi}\ppsi^\dagger\cchi +
	      \lambda_{\psi\chi}^*\cchi^\dagger\psi\ ,
\labeq{lagrangian}
\ea
where the $\lambda$s are all Lagrange multipliers, with $\lambda_\psi,
\lambda_\chi$ real, and $\lambda_{\psi\chi}$ complex valued.  Note that the 
theory has one free parameter, the ratio of VEVs of the
two Higgs fields, $v_1/v_2$.
Using this
Lagrangian, we obtain classical field equations:
\ba
\del_\mu\del^\mu\ppsi + (\del_\mu\ppsi^\dagger\del^\mu\ppsi)\ppsi + 
{2v_2^2\over v_1^2 + v_2^2}(\del_\mu
       \cchi^\dagger\del^\mu\ppsi)\cchi & = & {\bf 0}\ , \nonumber \\
\del_\mu\del^\mu\cchi + (\del_\mu\cchi^\dagger\del^\mu\cchi)\cchi + 
{2v_1^2\over v_1^2 + v_2^2}(\del_\mu
       \ppsi^\dagger\del^\mu\cchi)\ppsi & = & {\bf 0}\ . 
\labeq{psichieqn}
\ea
(Compare equations (63) of~\cite{famsymm}.) 

For cosmological purposes, we are interested in evolving very inhomogeneous
field configurations, in regions where
a texture is collapsing and the
nonlinear terms in the equations are very important. There is
little hope of making much progress analytically in general, 
and recourse to numerical techniques is needed. 

We wish 
to discretize the field equations on a grid.  However, any direct
discretization of the field equations~\eqn{psichieqn} will fail immediately
and drastically: adding small corrections $\Delta\ppsi, \Delta\cchi$ according
to a discrete version of~\eqn{psichieqn} in general violates 
the
sigma model constraints $|\ppsi|^2 = |\cchi|^2 =1$, $\ppsi^\dagger\cchi = 0$.
With those gone, the fields will no longer represent a member of SU(3), and
the simulation becomes meaningless.

Interestingly, the standard method of dealing with this problem: to discretize
the action and enforce the necessary constraints with a discrete set of 
Lagrange multipliers, does not work here.  This is a pity, because this
technique works beautifully for SU(2) (see Pen et al,~\cite{bigso3}, eqns (67)
and (68)) and yields a very fast, efficient algorithm for evolving the SU(2)
field forward.  This method fails here because one must 
simultaneously solve for four Lagrange multipliers, ($\lambda_\psi,
\lambda_\chi$, and the real and imaginary components of $\lambda_{\psi\chi}$),
at each spacetime point. One needs to 
solve four coupled nonlinear (quadratic) equations
for four unknowns, at each point on the grid, at each timestep.  
(For SU(2), one only needs to solve single a quadratic equation.) This is 
numerically costly and complex - it must be done by iterative methods, 
and one has to worry about choosing the right one of the sixteen roots.

In order to solve the SU(3) field equations numerically, it is helpful 
to restate the problem 
in a form which more obviously respects the structure of the
group.  As written in~\eqn{psichieqn}, the field equations describe the motion
of two three dimensional complex scalar fields.  Thus there are twelve real 
degrees of freedom, which are four too many for describing a member of SU(3).
We have to try to control the remaining four dimensions by imposing 
constraints in some clever fashion.
It would be far better to work with a formulation of the field equations which
only included the eight real degrees of freedom that exist in the problem.

Note that this problem of extra degrees of freedom has nothing to do with
this being a field equation: the same problem would exist if we had only a
single pair of complex 3-vectors, which were constrained to be of unit length
and always perpendicular to one another.  In fact, this is analogous to a very
simple classical mechanical system.  If the two vectors were {\it real}, 
instead of complex valued, then the single point system would have the same 
dynamics as a pair of iron bars, of unequal length, welded together
perpendicular to one another, and with their point of contact held fixed, but
otherwise free to rotate, as in figure \fig{mech}.  In this analogous
system, the motion is trivial:  there are three degrees of freedom, so the
motion of this system is completely determined by conservation of angular
momentum.

\begin{figure}[th]
\centerline{\epsfbox{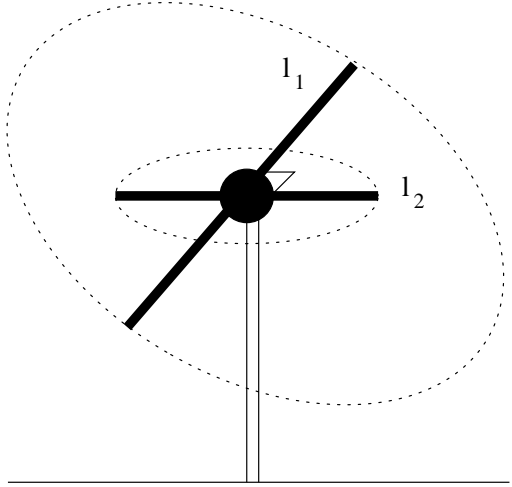}}
\caption{An analogous simple mechanical system.}
\labfig{mech}
\end{figure}

The pair of bars conserves angular momentum because their Lagrangian is
invariant under rotations in 3-space, SO(3).  Going back to the SU(3) breaking
Higgs field, the Lagrangian is clearly unchanged by an analogous global SU(3)
rotation, $\ppsi\rightarrow U\ppsi, \cchi\rightarrow U\cchi$.  This invariance
instantly gives us eight conserved Noether charges, components of angular
momentum on this internal SU(3) space.  Since our fields only have eight real
degrees of freedom, their dynamics are entirely determined by the conservation
of these Noether currents.  Further, if we formulate the problem entirely
in terms of the local member of SU(3), $U(x)$, and the local internal
angular momentum densities, then it will be written entirely in terms of
the eight real field degrees of freedom, and the corresponding velocities,
with no extra degrees of freedom to be dealt with.

In terms of Noether currents, then, the field equations are
\be
\del^\mu J^a_\mu = 0\ ,
\labeq{formaleqn}\ee
where the Noether currents $J^a_\mu $ are given by
\be
J^a_\mu = {\rm Im} \left\{v_1^2\ppsi^\dagger T^a\d_\mu\ppsi + {}
                         v_2^2\cchi^\dagger T^a\d_\mu\cchi \right\}.
\labeq{defJ}
\ee
Here the $T^a$ are the generators of SU(3); for sake of convenience we will
always use a basis where ${\rm Tr}[T^aT^b]= \delta^{ab}$.  Here and below,
we will use the convention that if $X^a$ are the components of a Lie 
algebra member, then $X = X^aT^a\in$ L(SU(3)) is the algebra element itself.

Now, define the local internal angular momentum densities, $L^a = J_0^a$.
In a flat FRW universe, the field evolution equations,~\eqn{formaleqn},
become
\ba
\d_\tau L^a &=& \d_iJ^a_i - 2 {\dot a\over a} L^a \nonumber \\
         &=& {\rm Im}\left\{v_1^2\ppsi^\dagger T^a\del^2\ppsi + {} 
                         v_2^2\cchi^\dagger T^a\del^2\cchi \right\}
			 - 2 {\dot a \over a} L^a\ .
\labeq{Levolution}\ea
Here $g_{\mu\nu} = a^2(\tau){\rm diag}\{1,-1,-1,-1\}$, so $a(\tau)$ is the
usual scale factor; we are working in 
conformal time and comoving spatial coordinates: 
we assume the background spacetime is flat Friedmann-Robertson-Walker.
$\del^2$ is the ordinary flat space Laplacian.  
The $\Im\{\d_i\ppsi^\dagger T^a\d_i\ppsi\}$ terms have vanished from 
\eqn{Levolution} because the hermiticity of $T^a$ makes the quantity in
brackets purely real.

With equation~\eqn{Levolution}, we have a straightforward, easily
calculable method for evolving the angular momenta, $L$, forward provided
we have the fields $\ppsi$ and $\cchi$.  In order to have a complete method
of evolving the fields forward, we need to be able to evolve the SU(3) field
forward, provided we know $L$.   To go back to 
the mechanical analogy: we need to solve for angular velocity in terms of 
position and angular momentum; we need a moment of inertia tensor.  Using the
condensed notation of~\eqn{Udef}, we need $\d_\tau U$.  Now, since $U(x)$ must
always remain upon the group, we know that there exists some 
$\Omega(x)\ \in$~L(SU(3)) such that
\be
\d_\tau U = i\Omega U\ ,
\labeq{Omegadef}
\ee
and we need to get $\Omega$ from $L$.  The 
necessary transformation can be accomplished by taking the definition of $L$:
\be
L^a = J_0^a = \Im \left\{v_1^2\ppsi^\dagger T^a\bigdot{\ppsi} + \ 
		v_2^2\cchi^\dagger T^a\bigdot{\cchi}\right\}\ ,
\ee
and constructing the quantities:
\be
\ppsi^\dagger L^a T^a\ppsi,\qquad \ppsi^\dagger L^a T^a\cchi,
\qquad \ppsi^\dagger L^a T^a(\ppsi^\dagger\times\cchi^\dagger)\ ,
\qquad \mbox{etc.}
\labeq{icompts}
\ee
These components may be evaluated, to solve for 
$\bigdot{\ppsi},\bigdot{\cchi}$,
by using the following general relation:
For any complex 3-vectors, $A,B,C$,
\be
\Im\[ A^\dagger T^a B\] T^a C = {1\over 2} i\[ \( B^\dagger C\) A -
\( A^\dagger C\) B + {1\over 3}\( B^\dagger A - A^\dagger B\) C\] \ ,
\labeq{c-baccab}
\ee
a kind of complex generalization of the BAC-CAB rule for double cross
products. Equation~\eqn{c-baccab} may be readily proven by writing 
$ A^\dagger T^aB = \Tr\[ T^a B A^\dagger\] $, and then using the various 
known properties of the $T^a$s.

Using~\eqn{c-baccab}, then, we can evaluate the nine quantities of 
\eqn{icompts}, and solve for $\Omega$.  If we temporarily rotate coordinates
so that $U =1$, then if we write
\be
   L = \(
   \begin{array}{ccc}
   - L_{\psi\psi} - L_{\chi\chi}&L_{\alpha\psi}&L_{\alpha\chi} \\
   L_{\alpha\psi}^*&L_{\psi\psi}&L_{\psi\chi}   \\
   L_{\alpha\chi}^*&L_{\psi\chi}^*&L_{\chi\chi}
   \end{array}
   \) \ ,
\ee
(where {\boldmath $\alpha$} is a shorthand for $\ppsi^\dagger\times
\cchi^\dagger$), then $\Omega$ is
\be
\Omega = 2\( \begin{array}{ccc}
-\Omega_{\psi\psi} -\Omega_{\chi\chi}& {1\over v_1^2}L_{\alpha\psi}
                          &{1\over v_2^2} L_{\alpha\chi} \\
{1\over v_1^2} L_{\alpha\psi}^* & {2 L_{\psi\psi} + L_{\chi\chi}\over v_1^2}
                         &\( {1\over v_1^2 + v_2^2}L_{\psi\chi}\) \\
{1\over v_2^2} L_{\alpha\chi}^* & \({1\over v_1^2 + v_2^2}\)L_{\psi\chi}^* &
                        {2 L_{\chi\chi} +L_{\psi\psi} \over v_2^2}
\end{array}\) \ . 
\labeq{inertiatransform}\ee
This can more be written more concisely as $\Omega^a = {\bf I}^a_{\ b}L^b$, 
where ${\bf I}^a_{\ b}$ is the (almost diagonal) moment of inertia tensor, whose
action is defined by~\eqn{inertiatransform}.  

Finally,~\eqn{inertiatransform} is only valid
when $U=1$.  In order to obtain the general relation, we must do and undo
a rotation to the $U=1$ frame:
\be
   \Omega(L,U) =  U\left\{ {\bf I}\circ\(U^\dagger L U\)\right\}U^\dagger\ .
\labeq{Omegais}\ee

And that completes the circle.  Taking~\eqn{Omegais},\eqn{inertiatransform},
\eqn{Omegadef}, and~\eqn{Levolution} together, we have the field equations
in a form where we have explicitly solved for the time derivatives of two 
fields, with no 
extra degrees of freedom.  The equations upon which we will base the numerical
field evolution are:
\ba
\bigdot U &=& i\Omega(L,U)U\ , \labeq{Udot}\\ 
\bigdot {L^a} & = &{\rm Im}\left\{v_1^2\ppsi^\dagger T^a\del^2\ppsi + 
                         v_2^2\cchi^\dagger T^a\del^2\cchi \right\}
			 - 2 {\dot a \over a} L^a\ , \labeq{Ldot}
\ea
where the precise form of $\Omega(L,U)$ is given by~\eqn{Omegais},\eqn
{inertiatransform}.

\section{The numerical algorithm}
\label{numalg}

Now that we have the field equations,~\eqn{Udot} and~\eqn{Ldot}, we can set
about constructing a field evolution algorithm based upon them.  Space and
time are discretized into grids.  Since $U$ and $L$ are a field and its time
variation, it is natural to place them on alternate half timesteps.  We need
to construct a finite difference technique for evolving each field forward,
if the other is known one half timestep in the future.

\subsection{Discrete evolution of $L$}

 One can
read the algorithm for evolving $L$ forward directly from~\eqn{Ldot}:
\ba
L^a_{n+1/2} &=&{a^2(\tau_{n-1/2})\over a^2(\tau_{n+1/2})} L^a_{n-1/2} 
      \nonumber \\
     & & \quad + \ {\Delta_\tau\over \Delta_x^2}\Im\biggl\{
     v_1^2\ppsi_n^\dagger T^a \( \ppsi^{i+1}_n + \ppsi^{i-1}_n + ...
     - 6 \ppsi^{i,j,k}_n\) \nonumber \\
     & & \qquad\qquad\qquad
   +\ v_2^2\cchi_n^\dagger T^a \Bigl( \cchi^{i+1}_n + \cchi^{i-1}_n + ...
     - 6 \cchi^{i,j,k}_n\Bigr)\biggr\}\ , 
\labeq{discreteL}
\ea
where the $...$ signify similar sums over nearest neighbors in the $j$ and
$k$ directions. Here and elsewhere, if spatial indices are omitted, the 
$(i,j,k)^{\rm th}$ point is implied.   The spatial derivatives piece of~\eqn{discreteL} is
clearly second order accurate, and the factor
$\({a^2(\tau_{n-1/2})\over a^2(\tau_{n+1/2})}\)$  takes the effect of the 
expansion of the universe into
account: angular momentum density generically redshifts away with 
$a^{-2}$ as the universe expands.  (That looks a little odd because we are 
in conformal time.  In ordinary time, the angular momentum density would decay 
as $a^{-3}$, and integration over space with the proper volume element would 
give you a genuinely constant Noether charge.) 
Also,~\eqn{discreteL} has the pleasant property
that it exactly conserves the total amount of each component of global angular
momentum on the grid, since the sum over the entire grid of all of the
discrete second spatial derivatives will be exactly zero  (periodic
boundary conditions are enforced).

And so we have eight Noether charges which are conserved in the continuum, and
which are exactly conserved by this scheme of discretely evolving $L$ forward. 
Since this is only an exact {\it global\/} conservation, it says little about 
the accuracy of this method. (The local conservation of the internal 
angular momenta will only be correct to second order in~$\Delta_\tau$, in 
general.)  However, it does push heavily towards numerical stability,
for it is more difficult for solutions of finite difference equations to
diverge if they are required to separately conserve eight independent 
quantities while doing so.

\subsection{Evolution equations for $U$}

And so the evolution of the angular momentum field is well in hand.  Next,
we turn to the evolution of the SU(3) field, $U$.  We would like to have a
discrete version of the continuum equation,~\eqn{Udot}.  The first requirement
is that the evolution preserve the fact that $U \in$ SU(3), and so we want
something like:
\be
U_{n+1} = e^{i\Delta_\tau\Omega_{n+1/2}}U_n\ ,
\labeq{naive}
\ee
which would be second order accurate in time, and would preserve SU(3) 
membership up to machine accuracy.  However, in calculating~\eqn{naive},
one immediately runs into a snag: the moment of inertia transformation
from $L\rightarrow\Omega$ explicitly depends upon $U$.  Thus, in order to 
calculate $\Omega_{n+1/2}$ we need a value for $U$ one half timestep into the
future.  So we need some kind of predictive method to get a value for 
$U_{n+1/2}$, at least first order accurate in~$\Delta_\tau$.

Now, one can immediately think of several ways to construct such a thing:
for example, one could approximate $U_{n+1/2} = \sqrt{U_nU_{n-1}^\dagger}\,U_n
$.  However, in this construction, as in all the rest, one is immediately
confronted with ambiguities: is that last construction better, or worse, than
taking $U_{n+1/2} = \sqrt{U_{n-1}^\dagger U_n}\,U_n$?  Or should one try
some kind of mean of the two?  And there are a myriad of other possibilities, 
using the exponentials of combinations of $L_{n-1/2}, L_{n+1/2}$, together with
field values.  These all have similar ordering ambiguities.   The central 
problem is that constructions of the form~\eqn{naive} are trying to use 
group multiplication, instead of addition, to advance the fields.  This is 
lovely, and keeps one on the group, but unfortunately in~SU(3) group 
multiplication doesn't commute.  There will inevitably be ambiguities about 
the order in which the multiplication is to be done.

Instead of working on a group, and using multiplication, we should really work
on a space in which addition is legal, and so no ambiguities of ordering can 
arise.  The obvious such space is the Lie algebra: we should take the logarithm
of the continuum equation~\eqn{Udot}, and discretize that instead.  If we
write
\be
U(x) \equiv e^{iF(x)}\ ,\qquad F(x)\in \mbox{L(SU(3)),}
\labeq{Fdef}
\ee
then~\eqn{Udot} becomes:
\be
\d_\tau e^{iF} = i\Omega\(L,e^{iF}\)e^{iF}\ .
\labeq{sub1}
\ee
Now we need to solve~\eqn{sub1} for $\bigdot F$.  This may be accomplished
by means of the following construction (well known in lattice gauge theory):
\begin{eqnarray*}
i\Omega\ e^{iF}  & = & \d_\tau e^{iF} \\
e^{-iF}i\Omega\	e^{iF}	& = & e^{-iF}\d_\tau e^{iF} \\
		& = & \int_0^1 \derv\lambda
	\(e^{-i\lambda F}\d_\tau e^{i\lambda F}\) \dee\lambda \\
		& = & \int_0^1
	\[ - iFe^{-i\lambda F}\d_\tau e^{i\lambda F}
	+ ie^{i\lambda F}\d_\tau\(F e^{i\lambda F}\) \] \dee\lambda \\
		& = & \int_0^1
	\[ - iFe^{-i\lambda F}\d_\tau e^{i\lambda F}
	+ ie^{-i\lambda F}\bigdot F e^{i\lambda F} 
	+ iFe^{-i\lambda F}\d_\tau e^{i\lambda F} \] \dee\lambda \\
		& = & \int_0^1 i e^{-i \lambda\Ad F}\ \circ\bigdot F\dee\lambda
		\qquad \mbox {by the definition of adjoint,}\\
		& = & i\[\int_0^1 e^{-i \lambda\Ad F}\dee\lambda\]
	\circ\bigdot F \\
		& = & i\({e^{-i\Ad F} - 1 \over -i \Ad F}\) \circ \bigdot F
\end{eqnarray*}
\ba
\bigdot F &=& { -i \Ad F \over e^{-i\Ad F} -1} e^{-i \Ad F}\circ \Omega
	\nonumber \\
	& = & {i\Ad F \over  e^{i\Ad F}-1}\circ\Omega\ .
\labeq{Fdot}
\ea
And so we have an explicit expression for $\bigdot F$ in terms of $\Omega$.
Now, evaluation of~\eqn{Fdot} is  little tricky, since the $8\times8$ matrix
$e^{i\Ad F} - 1$ is singular, and~\eqn{Fdot} is written in terms of its 
inverse.  There is no real problem, since the $\Ad F$ on top is also singular,
and the two singularities cancel one another out.  However, one cannot
calculate the right hand side of~\eqn{Fdot} directly; one must interpret
$\({i\Ad F\over e^{i\Ad F} -1}\)$ as the appropriate power series in~$i\Ad F$. 
So long as the series converges, there is no ambiguity as to the meaning of
\eqn{Fdot}.

We therefore have an expression for $\bigdot F$, which can be expressed in
terms of a power series we can find\footnote{Actually, the power series is
very well known, since ${x\over e^x-1}$ is a generating function for the
Bernoulli numbers.},  of quantities that we know.  At this point we could
use a discrete version of~\eqn{Fdot}, expressed as a power series, and attempt
to construct a field evolution algorithm based upon that.  That would be a
sound way to proceed, but slow, as the power series may take many terms to
converge.  (Indeed, sometimes it {\it doesn't\/} converge, as we discuss
in appendix A.)  Instead, with a little bit more maneuvering, we can construct
an exactly calculable expression for the right hand side of~\eqn{Fdot}.

We proceed as follows:  Write out the power series expansion of~\eqn{Fdot},
\ba
\bigdot F &=& \[1 + B_1i\Ad F + {B_2\over 2!}\(i\Ad F\)^2
+ {B_3\over 3!}\(i\Ad F\)^3 + ...\]\circ\Omega \\
	  &=& \Omega + B_1 i[F,\Omega] + {B_2 i^2\over 2!}\[F,[F,\Omega]\]
+ {B_3 i^3\over 3!}\[F,\[F,[F,\Omega]\]\] + ...\ \ ,
\labeq{series1}
\ea
by the definition of adjoint.  Here the $B_n$ are the Bernoulli numbers.  Now,
diagonalize $F$,
\be
F = RF_DR^{-1}\ ,\qquad\mbox{ and define } \Omega' = R^{-1}\Omega R\ .
\labeq{FOpdef}
\ee
With these terms, the power series~\eqn{series1} becomes:
\ba
\bigdot F &=& R\Omega'R^{-1} + B_1 i [RF_DR^{-1},R\Omega'R^{-1}] 
\nonumber \\ & & \qquad +
{i^2 B_2\over 2!}\[RF_DR^{-1},[RF_DR^{-1},R\Omega'R^{-1}] \] + ...
  \nonumber \\
   & = & R\biggl\{ \Omega' + i B_1 [F_D,\Omega'] 
   + {i^2 B_2\over 2!}\[F_D,[F_D,\Omega']\] + ...
   \biggr\} R^{-1}
\labeq{series2}
\ea
Now, we can express $\Omega'$ in terms of complexified Lie algebra elements
(step operators):
\be
\Omega' = \Omega'_D +\( \Omega'_T T_+ 
    + \Omega'_U U_+  +
   \Omega'_VV_+ + {\rm h.c.}\)\ .
\ee
(The notation used here is standard for SU(3), see e.g. ref.
\cite{chengli} pp.\,97--102.  Also, we abbreviate 
$\Omega'_D = \Omega'_{T3}T_3 + \Omega'_Y Y$ is the diagonal part of 
$\Omega'$.  The components of $\Omega'_D$ are purely real; the rest are 
complex.)  Now, in the complexified basis of the 
Lie algebra, the commutator of a diagonal matrix with any basis element is
proportional to that basis element.  Thus we have a chain of identities
of the form
\be
\begin{array}{rcr}
\mbox{$[\underbrace {F_D,[F_D, ... [F_D}_{n\ {\rm terms}} ,T_+ ] ]...]$}
               & =& \alpha_{{}_{T+}}^n T_+ \ ,
\end{array}
\ee

\noindent
and similarly for all of the other basis elements. Here $\alpha_{{}_{T+}}(F_D)$ is
a simple real valued function of the diagonal components of $F_D$. 
Using these relations, we can write  the power series as
\def\efunc#1{{i\alpha_{{}_{#1}}\over e^{i\alpha_{{}_{#1}}}-1}}
\ba
\bigdot F &=& R\biggl\{\(1+B_1 i \alpha_{{}_{T+}} 
   + {B_2\over 2!}\(i\alpha_{{}_{T+}}\)^2  + ...
   \)\Omega'_{T+}T_+ \nonumber \\
          & &\qquad\qquad +\ \mbox{ similar terms in other basis elements }
   \biggr\}R^{-1} \ .
\ea
And now we can explicitly resum the power series, which is now just
a power series in an imaginary number.  This gives
\ba
\bigdot F &=& R\biggr\{ \Omega'_D + 
   \efunc{T+}\Omega'_TT_++ \efunc{V+}\Omega'_VV_+   \nonumber \\
	  & & \mbox{\hskip 1.5 in}  + \efunc{U+}\Omega'_UU_+ 
	  + {\rm h.c.s}\  \biggr\}R^{-1}\ . \labeq{FinalFdot}
\ea
\newcounter{ffd}   
\setceqc{ffd}{equation}   
The diagonal components of $\Omega'$ are uneffected because they commute
with the diagonal matrix $F_D$, so $\alpha_{{}_T3}=\alpha_{{}_Y}=0$.  All other components
of the complexified basis of the Lie algebra have some nontrivial commutator
with the $F_D$, and so pick up an extra factor of $\efunc{{}}$ by this 
transformation.  However, the $\alpha\,$s are just real numbers, and so
now this function is very easy and fast to calculate.

And so we are done:  in~\eqn{FinalFdot} (together with definitions in 
\eqn{FOpdef},\eqn{Omegadef})  $\bigdot F$ is expressed entirely in
terms of elementary functions of $F,\ U= e^{iF}$, and $L$.  The only fancy
operation necessary is the diagonalization of a $3\times 3$ hermitian matrix.
Everything else is addition, multiplication of $3\times 3$ complex matricies, and exponentiation and arithmetic of complex numbers.  

Everything in~\eqn{FinalFdot} can be calculated explicitly, and to
machine accuracy, without resorting to iteration.  And, since $F$ has
by construction precisely the eight physical degrees of freedom that
exist in the problem, evolution of $F$ is guaranteed to keep you
exactly on the group.  Finally, since $F$ is a member of an algebra,
addition is legal, and so we may in fact justly
discretize~\eqn{FinalFdot} as a difference equation.

There is only
one more sensitive point which has not yet been addressed: the Lie
algebra has a different topology than SU(3). Thus, one should be quite 
suspicious of any method of studying topological defects in~SU(3) which looks 
at the group through the Lie algebra; the exponential map between the algebra 
and the group is not one to one.  This point is explored in appendix~A; the result is that one must be a little careful to always apply 
 equation~\eqn{FinalFdot} in a context in which it is meaningful.  However, 
this does not materially change change the field evolution scheme.  Also,
note that the evolution equation for $L$,~\eqn{discreteL}, depends only upon
the SU(3) group members, $U \ (= [..,\ppsi,\cchi])$, and not upon the Lie 
algebra member $F$.  So there will be no problems with the evolution of the 
$L$ field coming from the topology of the Lie algebra attached to $F$.

\subsection{The finite difference algorithm}

And so, we are finally in a position to construct a finite difference
algorithm for evolving the SU(3) nonlinear sigma equations.  
The differencing 
scheme we use for $L$ has been given; for $F$ we need to use some scheme
to predict $F$ one half timestep into the future, since $\bigdot F$ depends 
upon $F$.  To make this prediction accurately, we use a ``$-{1\over 4},
+{3\over 4},+1$'' differencing scheme as follows.     If we abbreviate
\eqn{FinalFdot} as
\be
\bigdot F = G(F,U,L)\ ,
\ee
then, assuming $F_n,F_{n-1},L_{n-1/2},L_{n+1/2}$ are known, we construct:
\ba
F_{n-1/4} &=& {3\over 4} F_n + {1\over 4} F_{n-1} \labeq{mfourth}\\
F_{n+1/2} &=& F_{n-1/4} + {3\over 4}\Delta_\tau 
            G\bigl(F_n,U_n,\mbox{${1\over 2}$}(L_{n+1/2} + L_{n-1/2})\bigr)
\labeq{Fhalf}\\
F_{n+1}   &=& F_n + \Delta_\tau G\(F_{n+1/2},U_{n+1/2},L_{n+1/2}\)\ .
\labeq{discreteF} 
\ea
And, again, the $L$ field is evolved by
$$ \begin{array}{rcl}
L^a_{n+1/2} &=&{a^2(\tau_{n-1/2})\over a^2(\tau_{n+1/2})} L^a_{n-1/2} 
      \nonumber \\
     & & \quad + \ {\Delta_\tau\over \Delta_x^2}\Im\biggl\{
     v_1^2\ppsi_n^\dagger T^a \( \ppsi^{i+1}_n + \ppsi^{i-1}_n + ...
     - 6 \ppsi^{i,j,k}_n\) \nonumber \\
     & & \qquad\qquad\qquad
   +\ v_2^2\cchi_n^\dagger T^a \Bigl( \cchi^{i+1}_n + \cchi^{i-1}_n + ...
     - 6 \cchi^{i,j,k}_n\Bigr)\biggr\}\ , 
\end{array}\eqno{\eqn{discreteL}}
$$
This difference scheme leads to an approximation to $F_{n+1/2}$ which is
second order accurate in~$\Delta_\tau$ if $L$ is changing slowly. The difference
scheme is shown geometrically in figure \fig{difference}; two second order, the
correct path between the points between the points $(F_{n-1},F_{n},F_{n+1/2})$
is a circle whose tangent is given by $\bigdot F_n$.  If angular velocity is 
changing reasonably slowly, the angle between $F_n$ and $F_{n+1/2}$ should be 
half the angle between $F_n$ and $F_{n-1}$, and a simple geometrical 
construction shows that a line parallel to the tangent at $F_n$ will cross 
$F_{n+1/2}$ if started from the point labeled $F_{n-1/4}$, as defined 
in~\eqn{mfourth}. In the worst case, when the angular velocity about the
circle is changing rapidly, this will be at worst $1^{\rm st}$ order accurate
in~$\Delta_\tau$, and so $F_{n+1}$ will always be at least second order accurate
in~$\Delta_\tau$.

\begin{figure}[ht]
\labfig{difference}
\centerline{\epsfbox{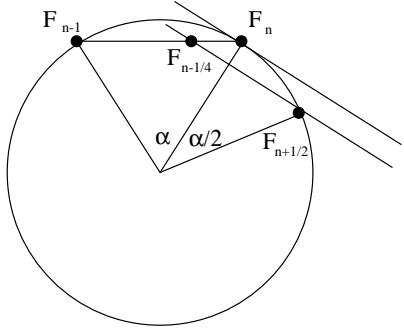}}
\caption{The differencing scheme.  To second order, we may approximate the
correct path as a circle defined by $F_n$, $F_{n-1}$, and the tangent line 
$\dot F_n$.  The line between $F_{n-1/4}$ and $F_{n+1/2}$ is parallel to the 
tangent at $F_n$.}
\end{figure}

Experimentally 
this works {\it much\/} better than any of the (faster) first order 
accurate schemes we used for constructing $F_{n+1/2}$.
This scheme is somewhat costly in computation time: it involves two
calculations of the complicated function $G(F,U,L)$ per timestep, on each
grid point.  However, this differencing scheme, combined with the discrete
$L$ evolution equation~\eqn{discreteL} (together the technicality discussed in 
appendix~A), yields an evolution algorithm which
is {\it extremely\/} stable (solutions do not diverge even when totally 
random initial conditions are given, in Minkowski space. Minkowski space is
the hardest case, since there is no damping coming from the expansion of the 
universe.) Further, it's behavior is very accurate by every test we have put 
to it.

\section{Testing the numerical code}
\label{testcode}

In the last section, we presented a finite differencing algorithm, 
equations~\eqn{discreteL}, \eqn{mfourth}--\eqn{discreteF}, designed
to approximate solutions to the field equations for a broken global SU(3)
symmetry, which were developed in section~\ref{formulation}.
We have subjected this algorithm to numerous tests of its accuracy and 
stability. This section summarizes these tests, and their outcomes. The central
conclusion is that the algorithm is stable, under any set of initial conditions
with $L_{\rm initial}=0$, provided the timestep ${\Delta_\tau\over\Delta_x}
\ \simlt\ 0.3$.  
Further, when the algorithm is stable, its accuracy
(as measured by violation of local conservation of energy) goes as 
$\({\Delta_\tau\over\Delta_x}\)^2$.  Also, when fed initial conditions where
the analytic solution is known, or where the field behavior may be compared
with the results of other programs, this program produces results consistent
with the correct or previously found ones.  Unless otherwise specified, all 
tests were done in Minkowski space, with $a=1,\ \dot a=0$.

\subsection{Stability}

Firstly: as to stability.  This algorithm has a strong tendency to be stable,
since by the way it was constructed it will exactly conserve the eight global
components of internal angular momentum.  This does not, of course 
{\it enforce\/} stability, for this still allows a momentum component
$L_a\rightarrow \infty$ in one region and $L_a\rightarrow -\infty$ in another
region, the energy of the system diverging while preserving the correct 
total momenta.  Experimentally, we find that the total energy of the system
tends to diverge roughly linearly in time when 
${\Delta_\tau\over\Delta_x} \ge 0.4$,
and that total energy remains bounded, doing some random walk about a mean,
for ${\Delta_\tau\over\Delta_x} \le 0.3$.  Suprisingly, the simulation remains
stable even when we are totally unfair to it, putting in initial conditions
which are random group elements at each grid point, with no damping coming
from the expansion of the universe.  (In this case one cannot really interpret
the system as an approximation to a field; rather it is an array of randomly
oriented SU(3) members connected by springs.)  All of these results come from
runs where the initial momenta~$L$ are zero.  If one puts in initial momentum,
together with an initial field configuration consistent with that momentum, 
the system behaves predictably: its energy is stable unless one makes the 
initial momenta~$L$ so large that 
$\bigdot F\Delta_\tau \ \simgt\  0.4$ on a 
fair fraction of the lattice.  (Putting in an initial field 
configuration~$F_0,F_1$ which is {\it not\/} consistent with the initial 
momenta yields very unstable behavior, as might be expected).

\subsection{Local energy conservation}

Nextly, as to accuracy.  The best generic method we know of to test the 
accuracy of solutions is to watch the violations of local conservation of
energy.  (This is a much stronger test than merely looking at global energy
conservation, where local errors tend to be washed away by averaging.)
This check is a little bit tricky to implement, for to use it one must
construct discrete  derivatives of the discrete energy momentum tensor 
$\d^\mu T_{0\mu}$ which center properly upon a single grid point, and then
check the extent to which
\be
\delta_0T_{00} - \delta_iT_{0i} = \epsilon
\labeq{heuristic}
\ee
differs from zero.  The particular discretization chosen for the components 
of~\eqn{heuristic} are written out in appendix B.  We find that, with
a timestep of $\Delta_\tau/\Delta_x= 0.1$, with fairly smooth initial conditions,
the rms violation of energy conservation is
\be
 {\<|\,\epsilon\,|\>\over \<T_{00}\>} \simeq .02\ .
\ee
where the average is over grid points.  So for smooth fields, average violation of local
energy conservation is about 2\% per grid point, for a timestep of .1 grid
units.  With totally random initial conditions, this went up to about 
4--5\% per grid point, per timestep.  We found that this error scaled roughly 
with $\Delta_\tau^2$, as it should, and that it did not systematically tend in 
one direction or another.  Thus on large grids, the global energy conservation
looks ridiculously good, as many small errors average away.  We found that
for a fixed length of evolution time, the error in global energy conservation 
went with $\Delta_\tau^2$, for every class of initial conditions we tried.
These results remain unchanged when we put in the expansion of the universe
and added the necessary extra term to the the energy conservation equation. 

\subsection{Comparison with known behavior}

Next, we checked the behavior of the simulation in certain known 
circumstances.  Firstly, we set the $U$ equal to a constant over space,
gave it a constant angular momentum density $L$.  Not suprisingly, the entire
$U$ field rolls peacefully around SU(3) in unison; this was almost the only 
thing that it could do, since the angular momentum field $L$ will never change
under these circumstances.  For a less trivial test, we set $\cchi=(1,0,0)$ 
everywhere on the grid, and gave the $\ppsi$ field some initial conditions
$\ppsi=(0,\alpha,\beta)$. In this case a perfect evolution would never excite 
$\cchi$; all motion would remain on the SU(2) subgroup defined by fixing
$\cchi$.  This is also one circumstance in which results from this code can
be compared with results from another code, namely that for SU(2),
based on
very different principles \cite{bigso3}.  We found, firstly, that
$\cchi$ is in fact very little disturbed: when we put in a single large
texture on the $\ppsi$ field, and let it collapse, the $\cchi$ field was
excited to about $10^{-3}$, on average, away from its starting point 
(by a time of 8 grid units). During the same period, the $\ppsi$ field ranged 
over its full extent, from $-1$ to $+1$ on the two components free to move, 
since the texture had collapsed by this time.

Finally, we constructed the same texture in the $\ppsi$ field of this code
(again leaving $\cchi$ fixed, so the field should be purely on an SU(2) 
subgroup) and on the previous SU(2) code.  We evolved the two codes forward,
with the same timestep, on the same sized grid; if both codes evolved 
perfectly, the single texture would collapse identically on both grids.
The results of the two runs are quite dramatic: the two codes behave almost 
identically.

  The case of worst disagreement between the two codes happens 
just as the texture unwinds, when spatial gradients are largest and the two 
different discretization schemes 
must show themselves against the grid.  This worst case difference is shown
in figure \fig{collapse}, which are plots of the kinetic energy
of the fields, on a two dimensional slice through the center of the simulation 
box.  (The grids were 32 points on a side, and the texture was placed in the
center of the box, with no initial kinetic energy.  The fields were evolved 
forward with a timestep of .1 grid units.)  The energy scales in the
two pictures are the same; all numbers are in grid units.  Note that this
discrepancy is only to be expected: {\it both\/} simulations will necessarily
be inaccurate around the center of a collapsing texture in the final instants
of its collapse.  The instant of texture collapse is, by definition, a time 
when neighboring grid points at the center of the texture lie upon extremely 
different places in the group. No finite difference approximation to a partial 
differential equation will accurately describe such a situation.

Once the textures have collapsed and the gradients in the fields are smaller,
both codes may be expected to be much more accurate.  Indeed, later
in the simulation, as the energy of the textures is
radiating outward in a spherical shell, the two simulations are virtually
identical, as shown in figure \fig{later}.  Thus when 
its initial conditions are restricted to an SU(2) subgroup of SU(3), the 
behavior of this SU(3) code is in excellent agreement with the SU(2) nonlinear
sigma model code which has been previously developed.

\begin{figure}
\centerline{\hbox{\epsfbox{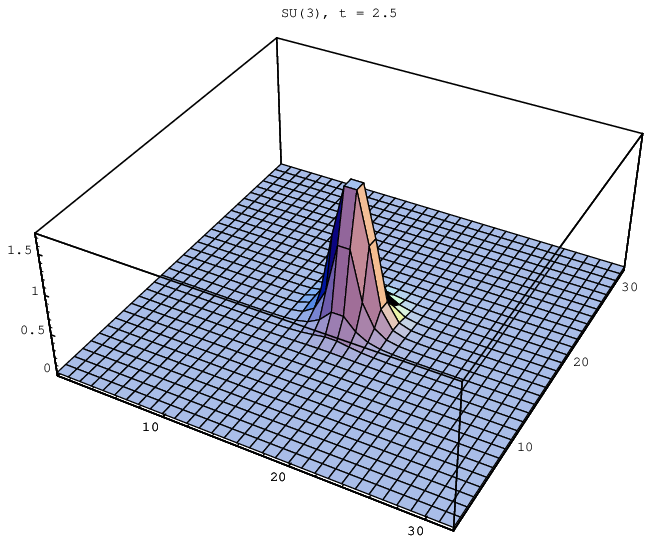}\hfil\epsfbox{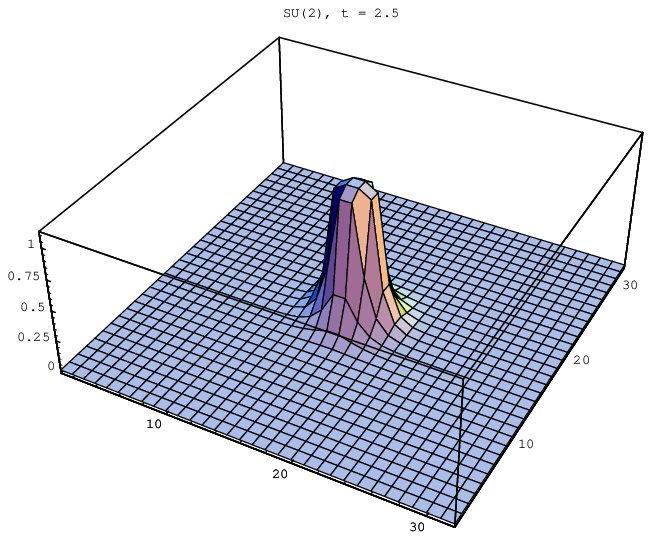}}}
\caption{The kinetic energy density on a two dimensional slice through the
center of a single texture, which was started with the same initial conditions
on the SU(3) and SU(2) codes.
This plot comes at the instant of texture collapse, when the field evolution
will be least accurate.}
\labfig{collapse}
\end{figure}

\begin{figure}
\centerline{\hbox{\epsfbox{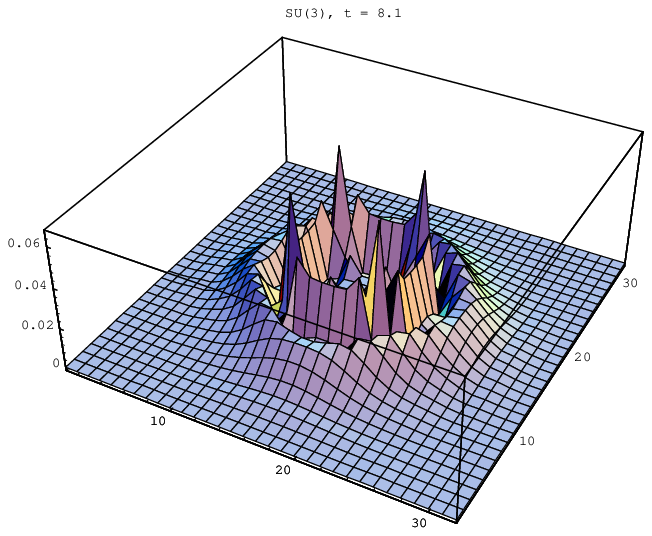}\hfil\epsfbox{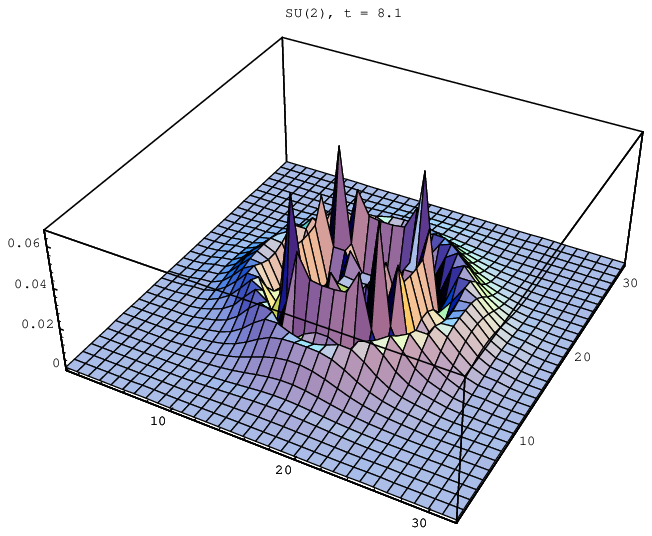}}}
\caption{Kinetic energy density on the same slice of the SU(3) and SU(2) 
simulation, after the texture has collapsed.}
\labfig{later}
\end{figure}

\section{Results}
\label{results}

This code was developed in order to study the effects of cosmic texture,
coming from a broken global SU(3) symmetry, upon the early universe.  In
particular, we would like to use this code to predict the appearance of
the cosmic microwave background, if the primordial density fluctuations
were seeded by this type of texture.  In order to make predictions about the
CMBR and density fluctuations, we need to evolve this field forward through
the early universe, calculating parts of its energy-momentum tensor as we
go: on the large scales of interest,
 this SU(3) field will interact only gravitationally with 
the matter and radiation of the early universe.  After calculating the
necessary parts of the energy-momentum tensor of this field, we can then
use this information to predict the power spectrum, nongaussianity, etc. of
the CMBR, using previously developed techniques.

In this section we describe the method of uniting this SU(3) evolution
code with other codes which evolve the matter and radiation of the
early universe, and give some results from these simulations.

\subsection{Scaling behavior}

In order to use this code to simulate the dynamics of an SU(3) field in the
early universe, we need to construct a set of initial conditions.  After the
initial quench, a physical SU(3) Higgs field would rapidly approach some
scaling solution, first in the radiation era and then in the matter era.  We
do not know, a priori, what the power spectrum of the scaling solution would
look like, and so we cannot put in realistic initial conditions.  However,
if we put in random initial conditions,  and evolve the field forward, the
simulation itself should also approach scaling behavior.  So the first thing
we must do is make sure that the SU(3) code does in fact approach a scaling
behavior.

In this case, we will look at the scaling behavior of one very
important quantity for the growth of gravitational perturbations, 
namely the source for Newtonian gravity, $\rho+3P \propto T_{00}+T_{ii}$. 
This acts as the gravitational source term in the
scalar density perturbation equations in synchronous gauge
(See e.g. \cite{twofluid_doppler}, \cite{durrer}).
This term alone is sufficient to determine the `intrinsic' perturbations 
to the CMBR (the synchronous gauge surface terms) and more 
generally, the matter perturbation power spectrum.
A brief
calculation shows that for this SU(3) system,
\be
T_{00}+T_{ii} = 2 \(v_1^2{\bigdot\ppsi}^2 + v_2^2{\bigdot\cchi}^2\)\ .
\ee
By scaling, we mean the scaling of the power spectrum of this quantity, 
\be
\<|\phi_k(\tau)|^2\>\ ,\qquad\mbox{where\ }\phi_k = {1\over(2\pi)^3 V}\int 
         \phi(x,t)e^{-ik\cdot x}\dee^3x
\ee
is the fourier transform of $\phi$.  If the simulation is purely in the 
radiation or matter era, there are no length scales in the system (save the
unphysical box size and grid size), so this power spectrum should reflect the
scale free nature of the problem.  Thus, in the radiation era, we require
\be
\<|\phi_k(\tau)|^2\> = {f(k\tau)\over \tau}
\labeq{scale_law}
\ee
for some function $f(k\tau)$.  (The extra factor of $\tau$ is necessary to balance
the units; $\phi(x)\sim {\bigdot \ppsi}^2$ has units of $\tau^{-2}$, and there is 
a factor of $\tau^3$ coming from the expansion of the universe).

Numerically, we take the average in \eqn{scale_law} by averaging over the
power spectra $\phi_k^2$ gotten from an ensemble of runs.  To test the scaling,
behavior we did an ensemble of six runs on $96^3$ grids in the radiation era,
starting from random initial conditions evenly distributed over SU(3).  As
can be seen from figure \fig{approachscale}, the power spectra in a very wide 
range of wavenumbers $k$ rapidly approach a scaling solution, and remain on 
it after $\tau\simeq6$ grid units.  

\begin{figure}
\centerline{\epsfbox{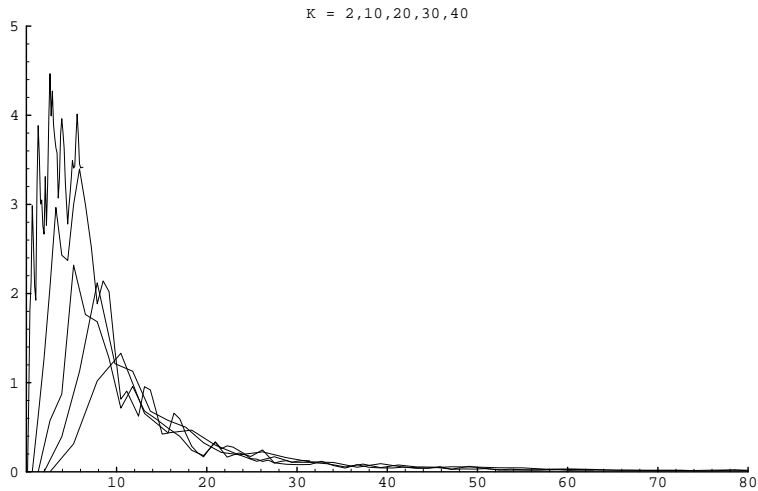}}
\caption{
The power spectrum of $\phi$ approaches scaling in the radiation era.  This is
a plot of $\tau \<|\phi_k(\tau)|^2\>$ vs. $k\tau$, each scaled to remove the box size.
Each trace in shows the time evolution of a different fourier mode;
each starts at zero because the simulation is started with the 
fields static, so $\phi(x,0) \sim \dot\psi^2 = 0$.  Each point on each trace 
was taken at a time $\tau$ one grid unit later than the preceding point; thus 
each trace starts at $\tau=0$, has its first joint at $\tau=1$, and so on.  
(Since the grid is 96 points on a side, the fourier wavenumbers 
run from k = (1--48)$\cdot{2\pi\over L}$.  Thus the range of $k$ shown covers 
almost all the wavelengths present in the simulation).  Each of these fourier 
modes represents an average of the power spectra from six separate runs.  
In each run, the power spectrum for a given $|k|$ is an average of power 
spectra in three perpendicular directions.  Thus each point here is an average 
of eighteen independent quantities.
}
\labfig{approachscale}
\end{figure}

Since scaling in this power spectrum of $\phi$ begins at $\tau\sim6$, we can find
the scaling function $f(k\tau)$ by averaging the values of $\tau\<|\phi_k(\tau)|^2\>$, 
for all $k$, for times greater than six grid units (and less than half the box
size, $\tau<48$ in this case.  For times later than this, the box size introduces
an unphysical time scale into the problem.)  The result of this averaging is
shown in \fig{avgscale}, which was constructed by summing all the values of
$\tau\<|\phi_k(\tau)|^2\>$ into bins in $k\tau$.  

\begin{figure}
\centerline{\epsfbox{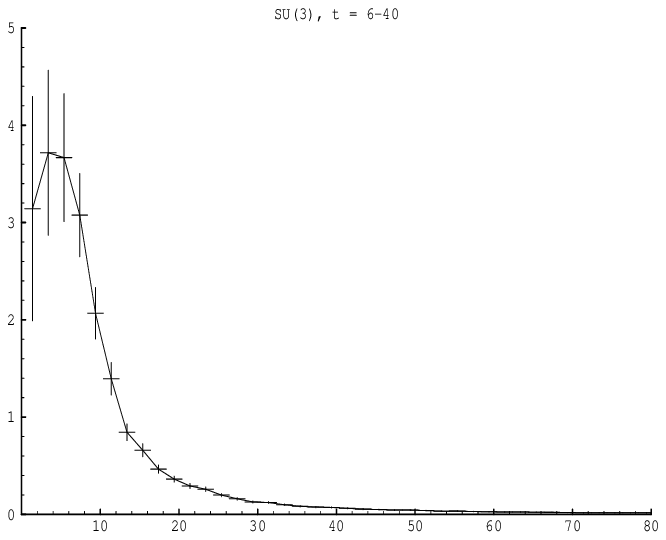}}
\caption{The averaged scaling function $f(k\tau)$ for SU(3) in the radiation epoch.
This is a plot of $f(k\tau)$ vs $k\tau$, and was constructed by averaging all values
of $\tau\<|\phi_k(\tau)|^2\>$ into bins in $k\tau$.  The average runs over all values of
$k$, for $\tau$ in a region where the power spectrum is scaling, $6<\tau<40$.
The vertical error bars come from assuming that $\phi_k(\tau),\phi_{k'}(\tau')$ 
will be independent from one another if $k\neq k'$.   Fourier components 
with the same
wavenumber, at different times, are not assumed to be independent.  The 
horizontal error bars are just the bin sizes.}
\labfig{avgscale}
\end{figure}

And with this scaling function, $f(k\tau)$, constructed, we are finally in a
position to begin comparing the behavior of SU(3) textures in the early 
universe with their SU(2) counterparts.  A similar set of six runs, on $96^3$
grids, in the radiation era, was done using the SU(2) code.  Using the same
analysis, we constructed a scaling function $f_{{}_{SU(2)}}(k\tau)$.  The two
functions are shown together in figure \fig{comparison}.  The resemblance is
striking: up to a multiplicative constant, the two 
functions are identical to within one another's error bars. The multiplicative 
constant is immaterial; it would be removed by normalizing both CMBR 
predictions to COBE, for example.  Thus, at least as far as one can
tell from looking at power spectra in the radiation era, SU(3) based
textures have almost exactly the same effect upon the early universe as
SU(2) based textures do.

\begin{figure}
\centerline{\epsfbox{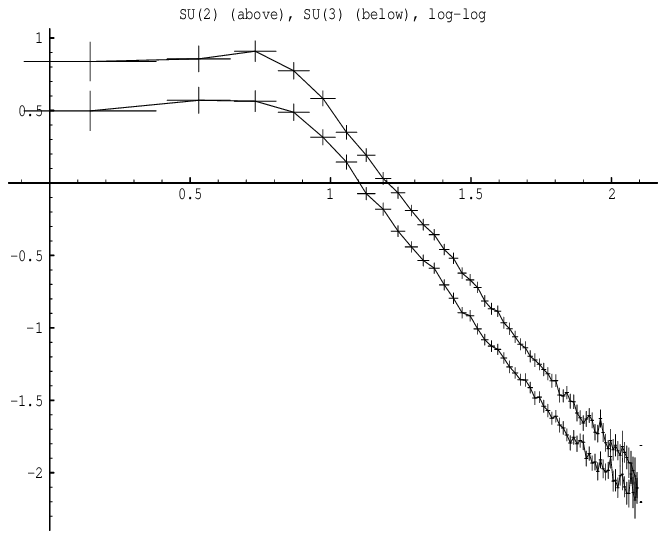}}
\caption{
 A comparison of the scaling functions $f_{{}_{SU(2)}}(k\tau)$ and
$f_{{}_{SU(3)}}(k\tau)$.  This is a plot of $\log_{10} f$ vs. $\log_{10} k\tau$
for both scaling functions; the SU(2) scaling function is the upper one.
Both of these scaling functions were created by averaging power spectra
from six radiation era runs on $96^3$ boxes.
}
\labfig{comparison}
\end{figure}

\subsection{CMBR predictions}

Finally, we used this code to provide the calculate the appearance of
the cosmic microwave background, in a universe where the large scale
structure was seeded by SU(3) textures.  In the simulations, we start
out the field with random initial conditions, and let the field evolve
forward in the radiation era long enough for scaling behaviour to set
in.  We then let the universe transfer from radiation to matter
dominated epochs, and continue forward in the matter epoch until the
time of last scattering.  During this simulation, we used the
energy-momentum tensor from the SU(3) field as a gravitational source
for the density and velocity fluctuations,
using code developed by Crittenden and one of us
(\cite{twofluid_doppler}, \cite{ntapjlett}).

From the density and velocity perturbations in the primordial gas of
one run, in a $256^3$ box, we made a set of 48 independant maps of the
visible temperature fluctuations on the surface of last scatter.  Two
such maps, which would appear as $10^\circ \times 10^\circ$ patches of
sky from earth, are shown in figure \fig{prettymaps}. Superficially,
the CMBR anisotropy maps presented here are very similar to previously
published textures maps. (See, for example, the pictures in
\cite{bigso3}.)  Note that these maps only represent the 
intrinsic anisotropies generated on the 
the surface of last scatter; they do not include the effects from
propagating the light from the surface of last scatter to us.  This
omission has little effect upon the small scale structure of the
image, but it does  mean that the
images have too little power on large angular scales ($l\ \simlt\ 200$).

\begin{figure}
\centerline{\epsfbox{su3a.eps}}

\vspace{0.5cm}

\centerline{\epsfbox{su3b.eps}}
\caption{Temperature anisotropies on the surface of last scattering. The
angular scale of these maps is $10^\circ$ on a side, for $\Omega =1$, 
$\Omega_B=0.05$,  and 
$h = .5$. The color scale units are in one standard deviation 
$\sigma$ for the CMBR anisotropy.}
\labfig{prettymaps}
\end{figure}

	From this assembly of images, one can readily calculate the
high end of the CMBR power spectrum, which is shown in figure
\fig{Cls} (along with similar plots from SU(2) textures, cosmic
strings, and monopole calculations).  Note that these maps don't
contain the necessary large scale information to normalize to COBE, so
the normalization of \fig{Cls} is left arbitrary. We do not expect
that the large scale behaviour of the CMBR spectra will be much
different between the SU(2) and SU(3) models, when it is calculated:
the interesting effects of the nonlinear dynamics show up on small
scales.  In particular, the positions and shapes of the doppler peaks
are correct in this plot.  Note that power spectum from the SU(3)
textures has a high, second peak, where the SU(2) power spectrum has
none, although the position of the first peak is the same.  So the
SU(3) field dynamics give a bit more power on smaller scales than does
SU(2).

\begin{figure}
\centerline{\epsfbox{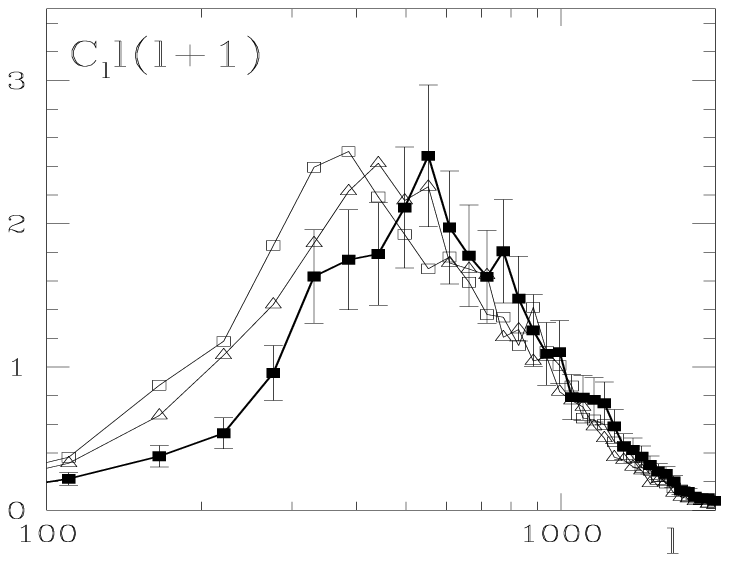}}
\caption{The region of the CMBR anisotropy power spectrum dominating 
on small angular scales for  
SU(3) textures (filled boxes), SU(2) textures (open boxes),
and global monopoles (open triangles).
All curves are for $\Omega =1$, $h = .5$. Error bars show the
statistical uncertainty (one sigma) in the SU(3) result.} 
\labfig{Cls}
\end{figure}

Finally, note that the CMBR maps of \fig{prettymaps} contain a handful
of extremely high peaks, many more than would be expected in a picture
of gaussian fluctuations.  These high peaks mark the infall of gas
into the gravitational wells created by the energy of the collapse of
individual textures: they are a primary source of primordial density
fluctuation in this theory.  The primary distinguishing features of
the textures CMBR maps is an overabundance of large ($\sim$5 sigma)
peaks, of which there are {\cal O}(1) in each map we created.  To
quantify this, we counted the number of peaks, of a given height,
which appear in a given area of sky in the collection of maps that we
created.  The distribution of peaks for SU(3) and SU(2) textures is
shown in \fig{peaks}.  Somewhat suprisingly, this measure of the
nongaussianity of the maps gives almost indistinguishable results for
SU(2) and SU(3) textures.  Thus the density of collapsing textures in
the universe is almost identical for the two theories.

\begin{figure}
\centerline{\epsfbox{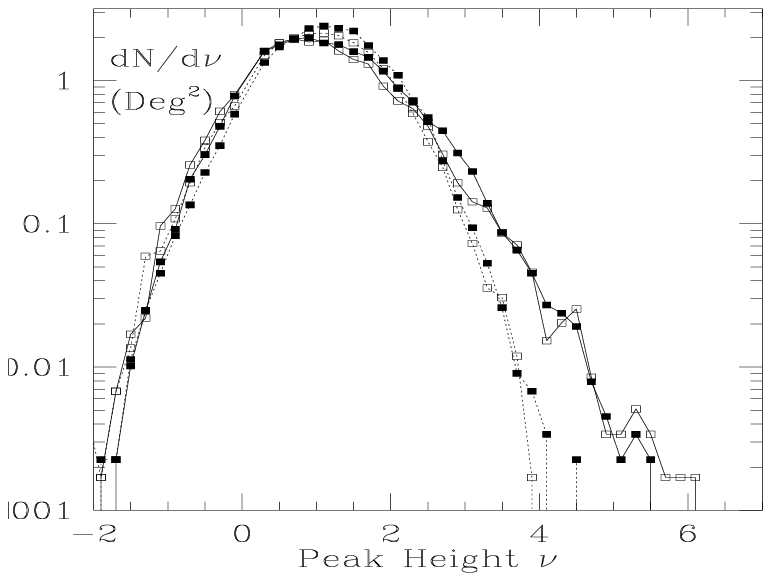}}
\caption{Differential number of local maxima (and minima) in the CMB 
anisotropy, of height $\nu \sigma$ (and $-\nu \sigma$) per
square degree, where
$\sigma$ is the rms. 
The number density of maxima is shown in solid lines, 
minima in dotted lines. The filled boxes show the SU(3) texture results, 
the open boxes the SU(2) results (from ref. [3]).}
\labfig{peaks}
\end{figure}

\section{Conclusion}

We have developed and used a reliable numerical technique for studying
the effects of textures from a broken global SU(3) symmetry upon the
early universe.  We find that, on the whole, SU(3)-generated textures
behave very similarly to textures which come from breaking a global
SU(2) symmetry.  In particular, the scaling behaviour of the two
theories in the radiation era, and the density of nongaussian peaks on
the cosmic microwave background, are identical in the two theories to
within our numerical errors.  We found that the CMBR power spectra of
SU(2) and SU(3) textures, are close, but do have a measurable
difference in shape: the SU(3) CMBR power spectrum has a bump not
present with SU(2) textures.  The two textures theories are much
closer to one another than they are to qualitatively different
theories of large scale structure formation.  This suggests that the
cosmological predictions of texture theories are rather robust:
changing the particular group which is broken to produce the textures
has relatively minor cosmological effects.  However with the
great accuracy one now anticipates in measurements of the
CMBR anisotropy power spectrum, even these small differences may be
subject to experimental test.

\section{Acknowledgements}  The computational power necessary for this
work was provided through the Pittsburgh Supercomputing Center grant \#AST9G3P.
We would also like to thank Guy Moore for helpful discussions on
computational techniques.

\begin{appendix}
\label{origins}
\section{The problems from nonuniqueness of $U\rightarrow F$}
   In section \ref{numalg}, we noted that there were some subtleties
in the evolution of the field $F$, the logarithm of the SU(3) field $U$, 
springing from the non-one-to-one nature of the map between the group
and its Lie algebra.  In this appendix, we discuss precisely how these
subtleties arise in the context of this numerical evolution, and how
they are handled.  We show that, if treated properly, this ambiguity never 
presents a problem in the field evolution or interpretation of results.

Recall that the continuum equation for the time derivative of $F$ was of
the form:
$$
\bigdot F = R\left\{\Omega'_D + \efunc{T+}\Omega'_{T+}T_+ +\ {\rm etc.}\
             \right\} R^{-1}\ ,
\eqno{\eqn{FinalFdot}}
$$
where $\Omega'$ was some Lie algebra element, $\Omega'_D$ is its diagonal
part, and the rest is split up in terms of the complexified basis.  Here 
$F$ has been diagonalized into $RF_DR^{-1}$, and the
$\alpha_{{}_{T+}}$, etc. are simple linear combinations of the components of
$F_D$.  In particular, if we break up $F_D$ into the usual generators of the
Cartan subalgebra on the main diagonal, $F_D = F_{T3} T_3 + F_Y Y$, then the $\alpha\,$s are:
\be
\alpha_{{}_{T+}} = F_{T3} \ ,\qquad\alpha_{{}_{U+}} = F_Y - {1\over 2}F_{T3} 
\ ,
\qquad\alpha_{{}_{V+}} = F_Y + {1\over 2} F_{T3}\ .
\ee
For all the generators, $\alpha_{{}_{X-}} = - \alpha_{{}_{X+}}$.

Now, the expression for $\bigdot F$,~\eqn{FinalFdot}, is written out in terms
of functions of the form $\efunc{{}}$ of the six $\alpha\,$s.  The behavior of
this function needs to be watched carefully, since our 
time derivatives are expressed in terms of it.  It has a removable singularity
at the $\alpha = 0$, which causes no trouble: the function smoothly approaches 
$i$  in that region.  However, the function has genuine singularities at 
$\pm 2 n \pi$, for all integers $n$ other than zero!  Thus if we 
na\"\i vely evolve the $F$ field forward according to~\eqn{FinalFdot} we rapidly
get into trouble: the time derivatives go infinite the first time one of the
$\alpha\,$s hits $\pm 2 \pi.$

And so direct application of~\eqn{FinalFdot} only makes sense when we are
within a certain neighborhood of the origin in the $F_{T3}, F_Y$ plane.
The situation is illustrated in figure \fig{hex}: the region of good behavior
of all $\alpha\,$s is bounded by a hexagonal wall of singularities of 
$\efunc{{}}$ in one or more $\alpha$.

\begin{figure}[ht]
\centerline{\epsfbox{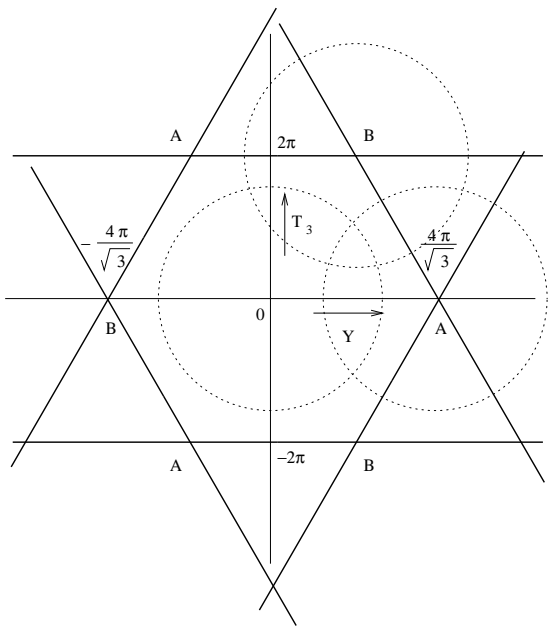}}
\caption {The singularities of equation~(\theffd), in the plane of
the $(T_3,Y)$ components of $F_D$.  The dark lines denote places where
some $\alpha = 2 \pi$, so $\alpha(e^{i\alpha} -1)^{-1}$ diverges.  The dotted
circles denote regions inside which $\alpha(e^{i\alpha}-1)^{-1} <{}$ 
some chosen bound, for all $\alpha$, for a particular choice of origin.}
\labfig{hex}
\end{figure}

Now these singularities are not physical, of course; the field equations
written in the form of~\eqn{Udot}, purely in terms of $U$, are not 
pathological anywhere on the group.  The singularities in~\eqn{FinalFdot}
are just symptoms of the fact that, when $F$ is far enough away from the 
origin, the coordinate system of the Lie algebra element $F$ is not a very 
good one for this type of calculation.

The solution to the problem is simply to change origins, whenever we get
far away from $F=0$.  That is, let us generalize our notation slightly,
so that the physical field $U$ is
\be
U = e^{iF}{\cal O}\ ,
\labeq{defO}\ee
where ${\cal O}$ is some origin, which we may choose to alter from time to
time as we advance the fields forward.  The generic effect of adding
an origin as in~\eqn{defO} is to change $F$ and its diagonalization:
\be
F_D \rightarrow F_D + \Delta_D\ ,\qquad R\rightarrow R'\ .
\labeq{Otransform}
\ee
We are permitted to change our origin in any way we please, and so we
can always choose the diagonal matrix $\Delta_D$ in such a way as to 
make $F_D$ smaller.  Provided we monitor the values of $F_D$, and change
origins whenever it approaches a singular region, we can always avoid
any difficulty.

In fact, figure \fig{hex} suggests a particularly nice set of origins to
use for this purpose.  We would like to choose the new origins to be places
which are as singular as possible in terms of the old coordinates, so as
to keep track of the smallest possible set of origins.  As such,
the points $A$ and $B$ on figure \fig{hex}, where two lines of singularities
intersect, naturally suggest themselves.  (The three points labeled $A$ on
figure \fig{hex} actually correspond to the same point in~SU(3), as do
the three points labeled $B$.  Every member of SU(3) can be made to correspond
to a point in the interior or on the boundary of the hexagon.  Within the 
hexagon, there is a three-fold ``reflection'' symmetry coming from the 
ambiguity of how we choose to order the eigenvalues of $F$ within~$F_D$). 
 In fact, the SU(3) members gotten by 
exponentiating $A$ and $B$ are just $e^{i\pi/3},\ e^{2i\pi/3}$ times the
identity: together with 1, they form the center of SU(3).  These points
are particularly nice origins because they make changing origin very easy;
they commute with everything.  In particular, the transformation of the
$R$ matrix in~\eqn{Otransform} becomes trivial, whereas in general it is a
painful thing to compute.

So: the evolution algorithm described in section \ref{numalg} is modified
as follows.  Each time we diagonalize $F$, we construct the radius${}^2$
$F_{T3}^2 + F_Y^2$ and figure out how far we are from our current
origin in figure \fig{hex}.  If we are outside of a certain radius, we
declare it time to change origins, and shift the origin of $F$ to the
corner of the hexagon nearest to the current position of $F_D$.  
(We must also shift the origin of
the previous timestep in the same way, because the algorithm uses the
previous timestep directly with this one.)  The only thing that need 
be decided systematically is which cutoff circle to use.  From figure 
\fig{hex} we see that the radius
must be greater than ${4\pi\over 3}$, so that the three circles centered at 
$1,A,B$ intersect,
and less than $2\pi$, to avoid the nearest singularity.  We chose 
$3\pi\over 2$, which provides a nice region of intersection, but is far
enough away from the nearest singularities that the time derivatives
never become too big; $(3\pi/2)|(e^{3i\pi/2}-1)^{-1}| \simeq 3.3$ is the
largest value obtained by the diverging function in that region. (The smallest
value obtained is 1, at the origin).  Now,
since the entire calculation of $\bigdot F$ will always be within some
circular region in which $\bigdot F$ is well behaved, we never encounter any
problems.

Everything else proceeds as before.  Of course, we
must now keep track of which of the three origins associated with each grid 
point, and make use of the origin as in~\eqn{defO} whenever we construct $U$.  
Fortunately, the simple form of the three origins chosen means that this does 
not slow down the program very much.

\section{Discretized derivatives of $T_{\mu\nu}$.}

In section \ref{testcode}, we describe the constructing the discrete
derivatives of the energy-momentum tensor, in order to test how well
the numerical algorithm obeys local conservation of energy, 
$\d^\mu T_{\mu0}=0.$  When discretized, this becomes
$$
\delta_0T_{00} - \delta_iT_{0i} = \epsilon\ ,
\eqno{\eqn{heuristic}}
$$
where $\epsilon$ is the error we seek to measure.
Now, \eqn{heuristic} involves a time derivative of the
00 component of $T$. A discrete version of $T_{00}$ lives most naturally upon 
a grid point on a whole timestep. The momentum densities $T_{0i}$ involve 
constructs like $\d_x\ppsi^\dagger\d_0\ppsi$, and so most naturally live
on the links between grid points, though on whole timesteps.  (We place
the time derivatives, $\bigdot\ppsi,\bigdot\cchi$ on whole timesteps
since we need to use both $U$ and $L$ to construct these.  The $L$ grid
lives on half timesteps, while $U$ lives on whole timesteps, so one or the
other will have to be averaged to produce a $\bigdot U$ at a single point.
Since $L$ lives in an algebra, averages of two $L$ are meaningful, whereas
na\"\i ve averages of two values of $U$ are not.  Thus we chose to construct
time derivatives which live on {\it whole\/} timesteps.)   Since 
equation~\eqn{heuristic} itself involves a time derivative of~$T_{00}$ and
spatial derivatives of the~$T_{0i}$, the discrete version of it may be fairly 
easily centered on a whole grid point, on a half timestep.  
\ba
\delta_0T_{00} &=& {1\over\Delta_\tau}\(T_{00}^{n+1} - T_{00}^n\)\ ,\qquad
                     \mbox{where}\\
T_{00}^n       &=& v_1^2\(\bigdot{\ppsi^\dagger_n}
		 \bigdot\ppsi_n\)  + {v_1^2\over 2\Delta_x^2}\Bigl(
	       \left|\ppsi^{i+1}_n - \ppsi^{i}_n\right|^2 +
	       \left|\ppsi^i_n - \ppsi^{i-1}_n\right|^2 \nonumber \\
	       & & \qquad\qquad + \mbox{ similar terms in $j$, $k$ directions}
	       \Bigr)\ ,
\ea
with a parallel term for $\cchi$.  Here the $\bigdot\ppsi$ terms are calculated
from the $L$ field, using the mean between the two half steps, as 
in~\eqn{Fhalf}.  Next we need to  construct
a discrete version of $\d_x T_{0x}$ which lives on a whole grid point, on
a half timestep.  Since the $T_{0x}$ live on whole timesteps, on links 
between lattice sites, this must be constructing by differencing adjacent
links, and averaging between two timesteps.  In one direction, this gives:
\ba
\left.\delta_xT_{0x}\right|^{n+1/2,i} &=& {1\over2\Delta_x}\biggl\{
      \(T^{n,i+1/2}_{0x} - T^{n,i-1/2}_{0x}\) + \nonumber \\
      & & \qquad\qquad
      \(T^{n+1,i+1/2}_{0x} - T^{n+1,i-1/2}_{0x}\)
			\biggr\}\ ,\qquad\mbox{where}\quad \\
T^{n,i+1/2}_{0x}  & = &{1\over 2\Delta_x}v_1^2\,{\rm Re}\left\{
\(\bigdot {\ppsi^i_n} + \bigdot{\ppsi^{i+1}_n}\)^\dagger\(\ppsi^{i+1}_n -
\ppsi^i_n\)
\right\}\ ,
\ea			
with, again, a similar term for $\cchi$.  

These are the explicit forms used for the tests of local energy conservation
used in section \ref{testcode}.  They are shown here in detail because there
several possible ways of constructing these discrete $T_{\mu\nu}$ components
and their discrete derivatives, and most of them do not work.

\end{appendix}

\end{document}